\documentclass[useAMS,usegraphicx,usenatbib]{mn2e}

\usepackage{graphicx}
\usepackage{times}
\usepackage{amssymb}
\usepackage{amsmath}
\newif\ifAMStwofonts
\AMStwofontstrue

\def\chandra{{\it Chandra}}
\def\xmm{{\it XMM-Newton}}

\def\ks{{\rm\thinspace ks}}
\def\ms{{\rm\thinspace Ms}}

\def\amsq{\hbox{$\rm\thinspace arcmin^{2}$}}
\def\degsq{\hbox{$\rm\thinspace deg^{2}$}}

\def\pdegsq{\hbox{$\rm\thinspace deg^{-2}$}}

\def\pcmsq{\hbox{$\rm\thinspace cm^{-2}$}}

\def\kev{{\rm\thinspace keV}}

\def\ctscmsqperg{\hbox{$\rm\thinspace count~cm^{2}~erg^{-1}$}}
\def\ergpcmsqps{\hbox{$\rm\thinspace erg~cm^{-2}~s^{-1}$}}
\def\ergps{\hbox{$\rm\thinspace erg~s^{-1}$}}

\def\kevpcmsqpspsrpkev{\hbox{$\rm\thinspace keV~cm^{-2}~s^{-1}~sr^{-1}~keV^{-1}$}}
\def\msunpmpccub{\hbox{$\rm\thinspace M_{\odot}~Mpc^{-3}$}}

\title[The unresolved hard X-ray background from \chandra\ and \xmm]
{The unresolved hard X-ray background: the missing source population implied by the \chandra\ and \xmm\ deep fields}
\author[M. A. Worsley et al.]
{\parbox[]{6.in}
{M.~A. Worsley,$^{1}$\thanks{E-mail: maw@ast.cam.ac.uk} A.~C. Fabian,$^{1}$ F.~E.~Bauer,$^{1}$
D.~M.~Alexander,$^{1}$ G.~Hasinger,$^{2}$ S.~Mateos,$^{3}$ H.~Brunner,$^{2}$
W.~N.~Brandt$^{4}$ and D.~P.~Schneider$^{4}$}\\\\
\footnotesize
$^{1}$Institute of Astronomy, Madingley Road, Cambridge CB3 0HA\\
$^{2}$Max-Planck-Institut f\"ur extraterrestrische Physik, Postfach 1319, D-85740 Garching, Germany\\
$^{3}$Instituto de F\'\i sica de Cantabria (CSIC-UC), 39005 Santander, Spain\\
$^{4}$Department of Astronomy and Astrophysics, 525 Davey Laboratory, Pennsylvania State University, University Park, PA 16802, USA\\
}

\begin{document}
\maketitle

\label{firstpage}

\begin{abstract} 
We extend our earlier work on X-ray source stacking in the deep \xmm\ observation of the Lockman Hole, to the \hbox{$2\ms$} \chandra\ Deep Field North and the \hbox{$1\ms$} \chandra\ Deep Field South. The \xmm\ work showed the resolved fraction of the X-ray background to be \hbox{$\sim80$-$100$} per cent at \hbox{$\lesssim2\kev$} but this decreased to only \hbox{$\sim50$} per cent above \hbox{$\sim8\kev$}. The CDF-N and CDF-S probe deeper, and are able to fill-in some of the missing fraction in the \hbox{$4$-$6\kev$} range, but the resolved fraction in the \hbox{$6$-$8\kev$} band remains only \hbox{$\sim60$} per cent, confirming the trend seen with \xmm. The missing X-ray background component has a spectral shape that is consistent with a population of highly obscured AGN at redshifts \hbox{$\sim0.5$--$1.5$} and with absorption column densities of \hbox{$\sim10^{23}$--$10^{24}\pcmsq$}.
\end{abstract}

\begin{keywords}
surveys -- galaxies: active -- cosmology: diffuse radiation -- X-rays: galaxies -- X-rays: diffuse background
\end{keywords}

\section{Introduction}

The nature of the X-ray background (XRB) has been a subject of contention since its discovery 40 years ago \citep{giacconi62}. It is now clear that that at energies above \hbox{$\sim1\kev$} the background is made up of the summed emission from point sources, principally Active Galactic Nuclei (AGN). The spectrum measured by the HEAO satellite \citep{marshall80} is similar in shape to \hbox{$40\kev$} bremsstrahlung and is well-described by a \hbox{$\Gamma=1.4$} power-law from \hbox{$\sim1$} up to \hbox{$\sim15\kev$} where it starts to flatten off before peaking in \hbox{$\nu I_{\nu}$} at \hbox{$\sim30\kev$}. Below \hbox{$\sim1\kev$} the background is dominated by diffuse Galactic and local bubble emission; the purely extragalactic component has not been well determined.

Resolving the XRB into discrete sources is crucial to understanding the nature of the AGN and galaxy populations in the Universe. It also allows us to place limits on the residual XRB emission which can be attributed to other processes, including that from truly diffuse extragalactic emission. A significant amount (\hbox{$\sim70$--$80$} per cent) of the \hbox{$0.5$--$2\kev$} XRB was resolved into point sources by ROSAT \citep{hasinger98} but progress at harder energies has required the improved sensitivity and spatial resolution of the \chandra\ and \xmm\ observatories. \xmm\ was able to resolve some \hbox{$\sim60$} per cent of the \hbox{$5$--$10\kev$} background in a deep survey of the Lockman Hole, reaching a limiting flux of \hbox{$\sim3\times10^{-15}\ergpcmsqps$} in this band \citep{hasinger01}. The \chandra\ Deep Fields North and South probe down to \hbox{$\sim1.4$--$2.8\times10^{-16}\ergpcmsqps$} (over \hbox{$2$--$8\kev$}) and resolve \hbox{$\sim70$--$90$} per cent of the background in this broad, hard band (\citealp{giacconi02,moretti02,alexander03}; and references therein).

The large resolved fractions found in the deep surveys appear to solve the problem of the origin of the XRB in terms of known sources; an extrapolation of the \hbox{$0.5$--$2\kev$} band \hbox{$\log N$--$\log S$} distribution points to an integrated flux consistent with the full XRB level. The substantial resolved fraction in the broad \hbox{$2$--$10\kev$} band is not entirely consistent with this picture; a similar extrapolation of the \hbox{$2$--$10\kev$} \hbox{$\log N$--$\log S$} distribution is, at most, only able to account for some \hbox{$93$} per cent of the total XRB and is only marginally consistent with complete resolution \citep{moretti03}. The most recent XRB intensity measurement by \citet{deluca04} would make the fraction only \hbox{$\sim80$} per cent.

To investigate in detail the resolved XRB fraction as a function of energy, \citet{worsley04} carried out a source-stacking analysis of \hbox{$\sim700\ks$} of accumulated \xmm\ exposure in the Lockman Hole. Source photometry was used to determine the resolved fraction in a number of narrow energy bands from \hbox{$0.2$} to \hbox{$12\kev$}. The authors found the resolved fraction to be \hbox{$\gtrsim80$} per cent below \hbox{$2\kev$} but that this decreased significantly at higher energies, falling to only \hbox{$\sim50$} per cent above \hbox{$\sim8\kev$}. The failure to account for XRB in the harder bands, along with recent indications of steepening of X-ray source number counts at low fluxes, suggest that there may be an as-yet undetected population of highly obscured sources, such as the heavily absorbed AGN predicted in recent synthesis models \citep[e.g.][]{gilli01,franceschini02,gandhi03,ueda03,comastri04}.

Although \chandra\ has low effective area at energies exceeding \hbox{$7\kev$}, the \chandra\ Deep Fields (CDFs) probe to fainter flux limits than the \xmm\ Lockman Hole observation. We use the CDFs to assess whether the contribution of the fainter sources found by \chandra\ are able to resolve the discrepancy and account for the missing background fraction. Using a similar photometric analysis we sum the flux from resolved CDF sources in a number of narrow energy bands to determine the resolved fraction of the X-ray background. We also present a re-analysis of the \xmm\ Lockman Hole observation \citep{worsley04}, carried out in identical energy bands for direct comparison with the CDFs. In both cases we attempt to quantify and correct for field-to-field variations and the missing flux which would arise from the very bright sources which are not sampled in these deep, pencil-beam surveys.

\begin{table*}
\centering
\caption{Parameters used in the XMM-LH photometric analysis. For each energy band the cut-out radii are quoted: $r_{1}$ is the radius of the source count-rate extraction aperture; $r_{2}$ and $r_{3}$ are the inner and out radii of the background count-rate extraction annulus; refer to sections 2.3-2.4 and Fig.~1 in Worsley et al. (2004) for further details. Also quoted are the count-rate to flux energy conversion factors calculated for each band following the method described in section 2.5 of that paper.}
\label{parameters}
\begin{tabular}{ccccccc r@{.}l r@{.}l r@{.}l}
\hline
Energy band & \multicolumn{6}{c}{Cut-out radii (arcsec)} & \multicolumn{6}{c}{Weighted energy conversion factors} \\
(keV) & \multicolumn{3}{c}{PN} & \multicolumn{3}{c}{MOS} & \multicolumn{6}{c}{($10^{11}\ctscmsqperg$)} \\
& $r_{1}$ & $r_{2}$ & $r_{3}$ & $r_{1}$ & $r_{2}$ & $r_{3}$ & \multicolumn{2}{c}{PN} & \multicolumn{2}{c}{MOS-1} & \multicolumn{2}{c}{MOS-2} \\
\hline
$0.2$--$0.5$ & $14$ & $30$  & $40$  & $12$  & $25$  & $35$  & 7&052  & 1&263   & 1&242 \\
$0.5$--$1$   & $15$ & $30$  & $40$  & $13$  & $25$  & $35$  & 7&375  & 1&757   & 1&735 \\
$1$--$2$     & $14$ & $30$  & $40$  & $12$  & $25$  & $35$  & 5&631  & 1&994   & 1&977 \\
$2$--$4$     & $11$ & $30$  & $40$  & $10$  & $25$  & $35$  & 1&960  & 0&7650  & 0&7664  \\
$4$--$6$     & $10$ & $25$  & $35$  & $8$   & $25$  & $35$  & 1&143  & 0&4017  & 0&4207  \\
$6$--$8$     & $8$  & $20$  & $30$  & $6$   & $20$  & $30$  & 0&6694 & 0&1347  & 0&1411  \\
$8$--$12$    & $6$  & $15$  & $25$  & $5$   & $15$  & $25$  & 0&1908 & 0&02059 & 0&02220 \\
\hline
\end{tabular}
\end{table*}

\section{Source detection and photometry}

\subsection{Chandra Deep Field Observations}

\chandra\ has performed the two deepest ever surveys of the X-ray universe -- the \hbox{$2\ms$} \chandra\ Deep Field North \citep[CDF-N;][]{alexander03} and the \hbox{$1\ms$} \chandra\ Deep Field South \citep[CDF-S;][]{giacconi02}. The CDF-N probes down to \hbox{$\sim2.5\times10^{-17}$} and \hbox{$\sim1.4\times10^{-16}\ergpcmsqps$} in the \hbox{$0.5$--$2$} and \hbox{$2$--$8\kev$} bands respectively. Although the CDF-S only reaches down to equivalent limits (at \hbox{$S/N=3$}) of \hbox{$\sim 5.2\times10^{-17}$} and \hbox{$\sim2.8\times10^{-16}\ergpcmsqps$}, these are both well beyond the \hbox{$10^{-14}\ergpcmsqps$} break in the \hbox{$\log N$--$\log S$} distribution \citep[see e.g.][]{hasinger98,campana01}.

For both the CDF-N and CDF-S we use the source catalogues and photometry from \citet{alexander03}, who describe their approach thoroughly; we include a brief summary here for completeness. The \hbox{$2\ms$} CDF-N main point-source catalogue contains a total of 503 sources and covers a total solid angle of \hbox{$447.8\amsq$}. Conservative estimates place the number of falsely detected sources to be as high as \hbox{$20$--$30$} but the true number is probably considerably lower (see section 2.3 in \citealp{alexander03}).  The CDF-S has 326 sources from a total field of \hbox{$391.3\amsq$}.

The local background estimate for each source was determined in an annulus surrounding the source-extraction region. The mean number of background counts was computed assuming a Poisson distribution, as \hbox{$n_{1}/n_{0}$}; where $n_{1}$ is the number of pixels containing a single count and $n_{0}$ is the number containing no counts. For sources with less than \hbox{$10^{3}$} counts in the full \hbox{$0.5$--$8\kev$} band, source-extraction apertures containing \hbox{$90$--$95$} per cent of the encircled energy fraction (EEF) were chosen (with the extracted counts corrected up to \hbox{$100$} per cent). For sources brighter than \hbox{$10^{3}$} counts, extraction apertures of twice this size were chosen; the EEF in these cases is close to \hbox{$100$} per cent. The EEF corrections used to correct from aperture counts to total counts were calculated from the \chandra\ point spread function (PSF) determined using the \chandra\ X-ray Centre \textsc{mkpsf} software (refer to \citealp{alexander03}). 

For all sources with a sufficient number of counts, the ratio of counts between the \hbox{$2$--$8\kev$} and \hbox{$0.5$--$2\kev$} bands was used to determine an effective spectral index for a power-law model, allowing for Galactic absorption of \hbox{$(1.3\pm0.4)\times10^{20}$} and \hbox{$(8.8\pm4.0)\times10^{19}\pcmsq$} for the CDF-N and CDF-S respectively \citep{lockman04,stark92}. For sources with an insufficient number of counts, the photon index is fixed at \hbox{$\Gamma=1.4$} (the approximate average for all the sources and the slope of the XRB). The flux for each source was calculated in a number of photometric bands, regardless of whether or not the source is actually detected in each of the bands. In each case the counts-to-flux conversion is appropriate to the spectral index of the source. The photometric bands used were \hbox{$0.5$--$1$}, \hbox{$1$--$2$}, \hbox{$2$--$4$}, \hbox{$4$--$6$} and \hbox{$6$--$8\kev$}.

Prior to our stacking analysis (described in section~\ref{resolving}), we applied the systematic flux correction determined by \citet{bauer04} for both the CDFs (see section 3 and Fig. 2 of their paper). These take into account Eddington bias as well as some further aperture/photometry effects that were not considered originally. The correction is negligible except in the harder (\hbox{$>2\kev$}) bands which each see an increase in total resolved flux of \hbox{$\sim2$-$5$} per cent.

\subsection{XMM-Newton Lockman Hole Observations}

\xmm, whilst unable to match the X-ray resolution and ultimate sensitivity of \chandra, has the critical advantage of sensitivity in the \hbox{$8$--$12\kev$} regime. It was this capability, put into effect with \hbox{$\sim700\ks$} of deep observations in the Lockman Hole (XMM-LH), that prompted \citet{worsley04} to investigate the resolved fraction of the XRB with energy.

The 17 different individual observations that comprise the XMM-LH were combined and source detection carried out using sliding-box detection followed by maximum-likelihood PSF fitting. Visual inspection of the source candidates was used to exclude spurious detections. The analysis was restricted to the central \hbox{$314.2\amsq$} region of the field with the highest exposure time and lowest background level. 

In a similar approach to that taken in the CDFs, simple aperture photometry was performed on the exposure-corrected images in a number of narrow energy bands. A more detailed description of the following is given by \citet{worsley04}. For each source, the total count-rates were extracted in both a circular aperture and a surrounding annulus. These were then used to calculate the background-corrected source count rate, taking into account corrections for the EEF. These corrections were calculated by integrating an analytical model of the \xmm\ PSF given in \xmm\ Science Operations Centre calibration documents (refer to \citealp{worsley04}).Count-rate extraction apertures/annuli were chosen to the maximise the signal-to-noise ratio.

Count-rate-to-flux conversion factors were computed assuming a fixed spectral slope of \hbox{$\Gamma=1.4$} and Galactic absorption with a column density of \hbox{$5\times10^{19}\pcmsq$} \citep{lockman86}. This is not as rigorous as the approach taken in the CDFs where counts-to-flux conversion factors (ECFs) appropriate to the spectral slope of the source were used, but the larger count rate errors for the \xmm\ sources prohibit this, and the assumption of a fixed $\Gamma=1.4$ is more reliable. Furthermore, the dependence of the ECFs on the spectral shape of a source is only weak because of the use of narrow energy bands. The ECFs of all but the \hbox{$0.2$--$0.5\kev$} band vary by \hbox{$\lesssim5$} per cent over the range \hbox{$\Gamma=0.9$--$1.9$} (the \hbox{$0.2$--$0.5\kev$} band has a \hbox{$\sim15$} per cent variation). The resultant uncertainty in the resolved total flux from the source population is small in comparison with other errors.

In order to allow easy comparison with the \chandra\ data, the XMM-LH analysis was repeated using the same energy bands as in the CDFs with additional bands covering \hbox{$0.2$--$0.5$} and \hbox{$8$--$12\kev$}. The re-processing of the data took advantage of the latest releases of the \xmm\ \textsc{science analysis system} (\textsc{sas}) \textsc{v6.0.0} and the X-ray spectral analysis package \textsc{xspec} \textsc{v11.3.1}. Additionally, rather than use the original source list (from \citealt{worsley04}), of 126 objects, we took advantage of the complete -- and more rigorously compiled -- Lockman Hole source catalogue of Mateos et al. (in prep.), which contains 156 sources in the same \hbox{$314.2\amsq$} region. Table~\ref{parameters} gives the values of various important parameters used in the re-processing of the data.

\section{Results}

\subsection{Resolving the X-ray background}
\label{resolving}

The total resolved flux in each energy band was calculated by summing the measured flux for each source. Every source is included regardless of whether or not it is detected in the band. This was done for the CDF-N, CDF-S and each of the three \xmm\ instruments in the XMM-LH. Appropriate corrections for Galactic absorption were made. These correction factors are \hbox{$15$} per cent in the XMM-LH \hbox{$0.2$--$0.5\kev$} band; \hbox{$2.6$}, \hbox{$7.2$} and \hbox{$4.8$} per cent in the \hbox{$0.5$--$1\kev$} band for the XMM-LH, CDF-N and CDF-S respectively; and \hbox{$\lesssim1$} per cent for all of the harder bands. 

In order to calculate the resolved intensity, the flux from each source must be divided by a solid angle on the sky. In the case of the XMM-LH the sensitivity is approximately constant over the region considered here and we simply take the solid angle to be a constant \hbox{$314.1\amsq$}. The CDFs, however, show a substantial increase in sensitivity towards the centre of the fields. The total solid angles of \hbox{$447.8$} and \hbox{$391.3\amsq$} are only applicable to the brightest sources, whereas fainter sources are only detectable over a fraction of the field and the actual solid angles are smaller. Fig.~19 in \citet{alexander03} shows the solid angle as a function of $0.5$-$8\kev$ source flux in the CDF-N. We use this function (and its equivalent for the CDF-S) to provide the appropriate solid angle when converting each source flux to an intensity. We impose a minimum effective area of \hbox{$10$} per cent of the total to avoid catastrophically magnifying the errors in the fluxes from the faintest sources.

\begin{figure}
\rotatebox{270}{
\resizebox{!}{\columnwidth}
{\includegraphics{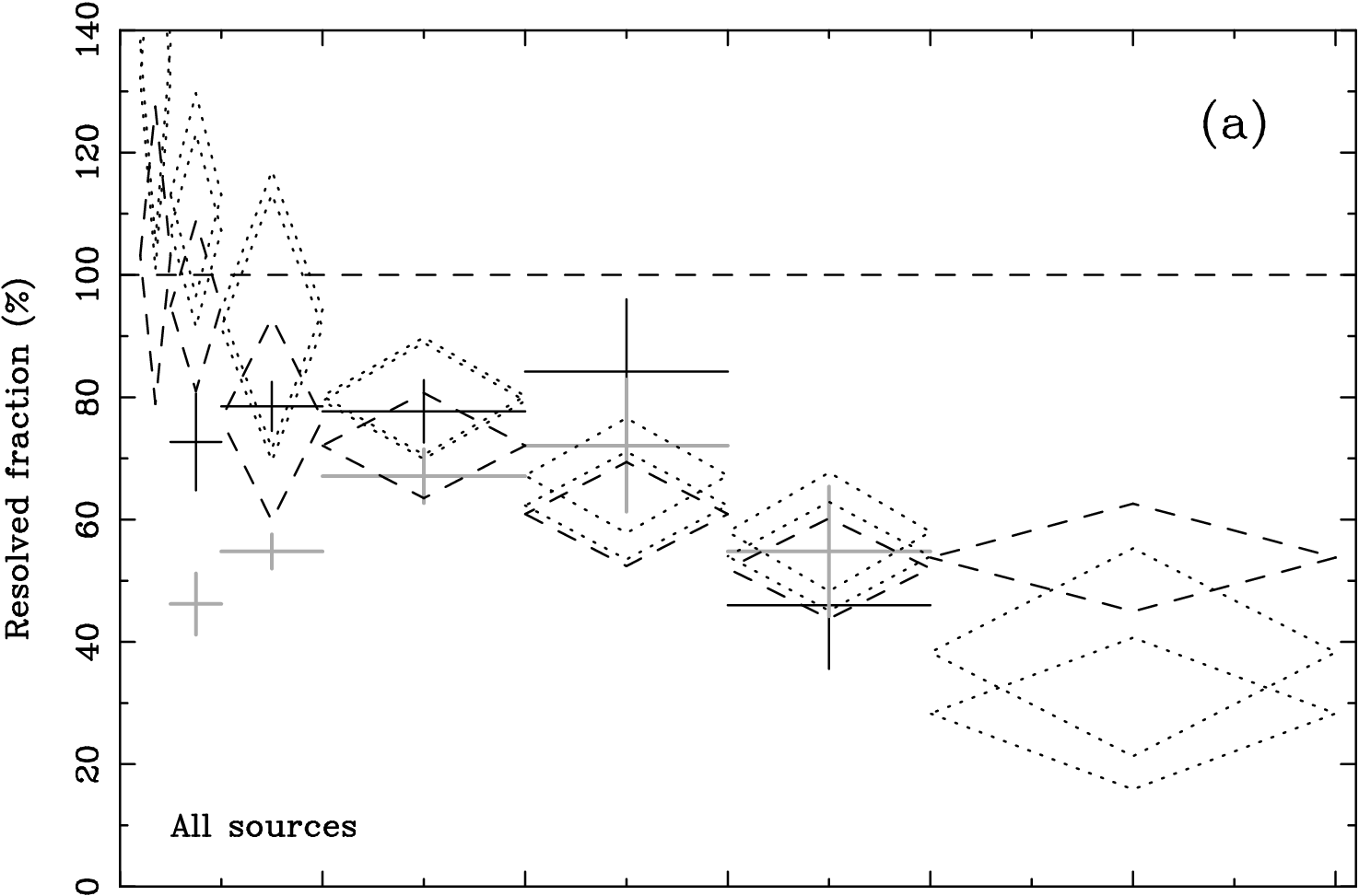}}
}
\rotatebox{270}{
\resizebox{!}{\columnwidth}
{\includegraphics{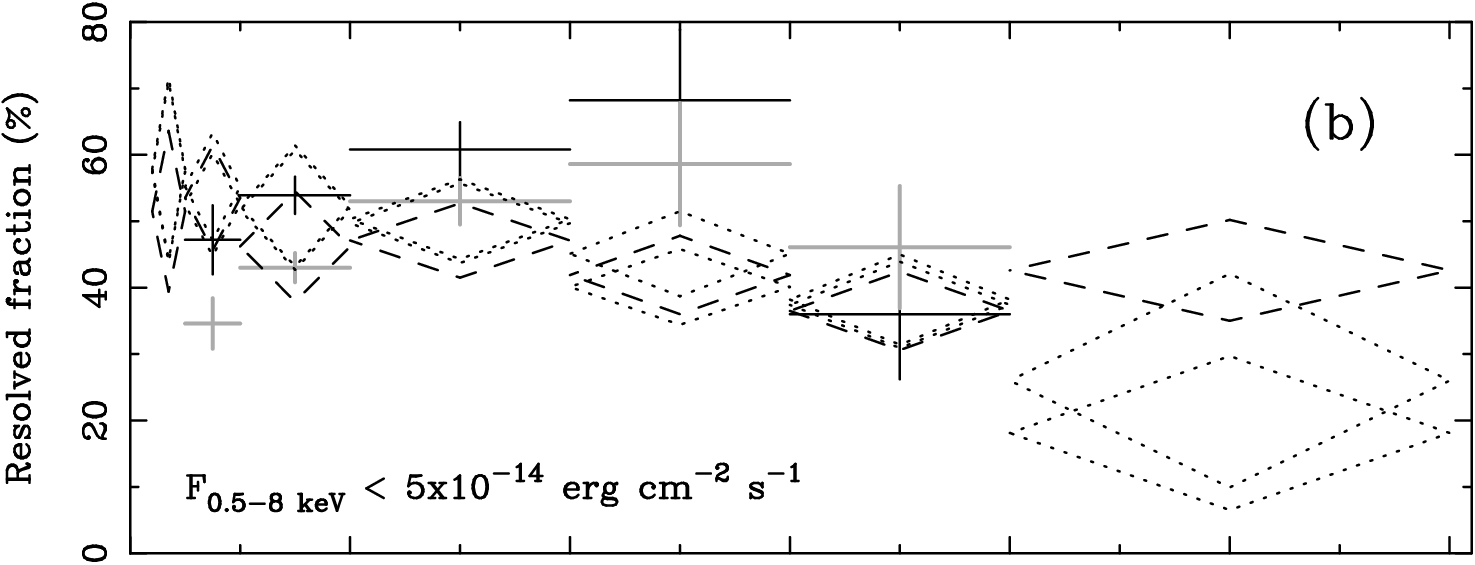}}
}
\rotatebox{270}{
\resizebox{!}{\columnwidth}
{\includegraphics{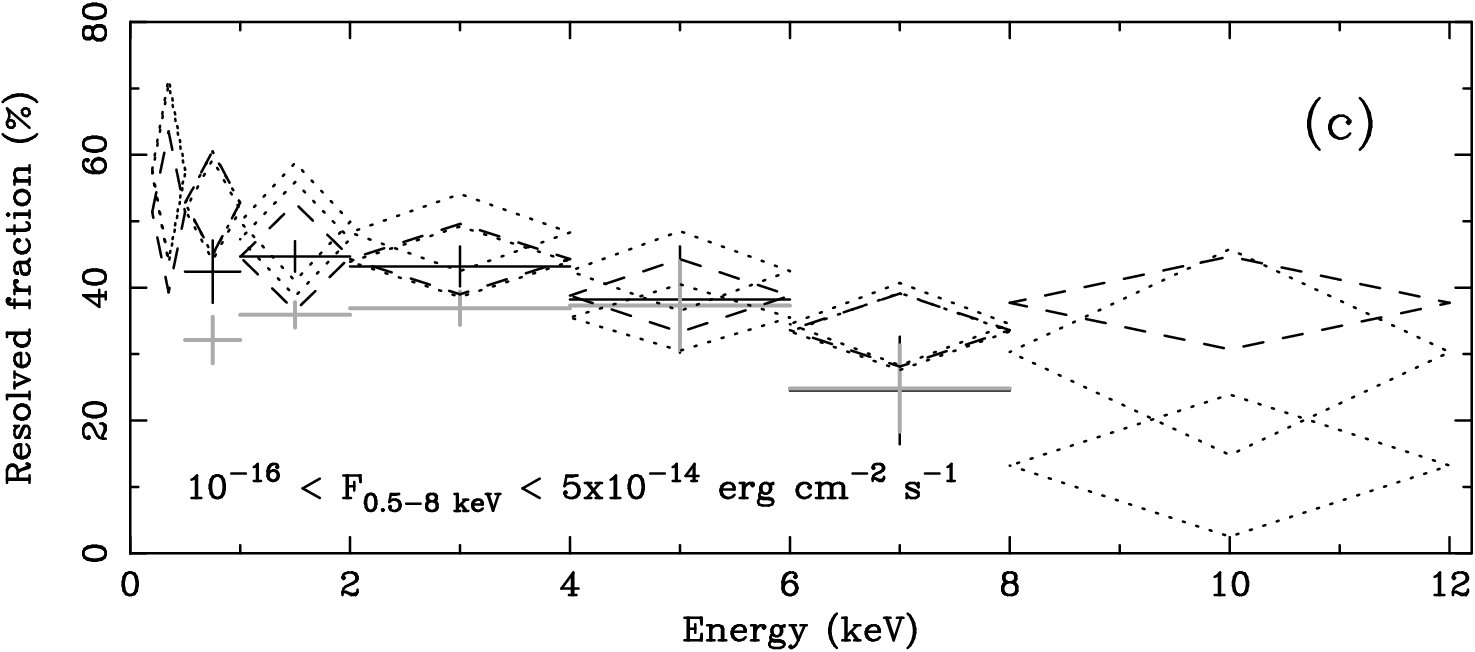}}
}
\caption{The fraction of the total extragalactic XRB intensity resolved by simply summing the fluxes of sources in the CDFs and XMM-LH. CDF-N and CDF-S data are shown as the black and grey crosses respectively; XMM-LH PN camera data are shown as dashed diamonds and both the MOS-1 and MOS-2 camera data are both shown using dotted diamonds. The three panels show; (a) the result of summing all sources, (b) the result of excluding sources brighter than \hbox{$5\times10^{-14}\ergpcmsqps$} in \hbox{$0.5$--$8\kev$} flux, and (c) the result of additionally excluding sources fainter than \hbox{$10^{-16}\ergpcmsqps$} in \hbox{$0.5$--$8\kev$} flux.}
\label{raw_fractions}
\end{figure}

Fig.~\ref{raw_fractions}(a) shows the resolved XRB fraction obtained in each energy band. The fractions are calculated using the most recent estimates of the total extragalactic XRB spectrum. For the \hbox{$1$--$8\kev$} band this is the power-law model of \citet{deluca04}, with a spectral slope of \hbox{$\Gamma=1.41$} and \hbox{$1\kev$} normalisation of \hbox{$11.6\kevpcmsqpspsrpkev$}. Above \hbox{$8\kev$} the background spectrum starts to turn over slightly and so here we use the analytical model of \citet{gruber99} -- which takes the downturn into account -- but renormalised to the same \hbox{$2$--$8\kev$} XRB flux as found by \citet{deluca04}. The spectrum of the purely extragalactic background is not well constrained below \hbox{$1\kev$} since it is difficult to separate from the bright, diffuse, Galactic component. There is evidence to suggest that the spectrum of the background steepens below \hbox{$\sim1\kev$}; \citet{roberts01} have combined the results of several X-ray shadowing measurements and report a background intensity of \hbox{$26.6\pm4.8\kevpcmsqpspsrpkev$} at \hbox{$0.25\kev$}. A power-law fit between this value and the \citet{deluca04} intensity at \hbox{$1\kev$} gives a spectral slope of \hbox{$\Gamma\sim1.5$--$1.7$} and we use this to estimate the background levels in the \hbox{$0.2$--$0.5$} and \hbox{$0.5$--$1\kev$} bands.

The stacking of source fluxes reveals a decline in the resolved XRB fraction as a function of energy. Below \hbox{$\sim2\kev$} the data diverge somewhat, presumably due to field-to-field variations. Such variations arise through the cosmic variance in the actual XRB normalisation in different regions of the sky, along with the statistical noise which is a consequence of sampling a discrete population of sources in a small solid angle. Fig.~\ref{raw_fractions}(b) shows the result of excluding sources with a \hbox{$0.5$--$8\kev$} band flux \hbox{$>5\times10^{-14}\ergpcmsqps$} from the sum. While only a small number of sources have been removed (from \hbox{$3$--$7$} out of \hbox{$156$--$503$}) the variation at the softest energies has been reduced considerably. This demonstrates that a large amount of the variation \hbox{$<2\kev$} is due to soft, bright sources. The XMM-LH data now show a slight decline in the resolved fraction whilst the CDFs show a slight rise, although both CDFs show significant drops in the \hbox{$6$--$8\kev$} band. This difference between the \xmm\ and \chandra\ fractions in the \hbox{$4$--$6\kev$} band is due to sensitivity -- the CDFs probe down to fainter fluxes where the typical source spectrum is harder than \hbox{$\Gamma=1.4$} \citep[see e.g.][]{alexander03,streblyanska03}. If the source lists are further restricted to remove the faint sources seen only by \chandra, with a minimum flux requirement of \hbox{$10^{-16}\ergpcmsqps$}, then the spectral shapes of the sources are similar for both XMM-LH and CDF, see Fig.~\ref{raw_fractions}(c). Of course, there will always remain differences between data-sets -- \citet{barcons00} estimate the cosmic variation in XRB normalisation to be of order \hbox{$\sim10$} per cent for fields of \hbox{$\lesssim1\degsq$}.

\subsection{Correcting for the bright-end population}
\label{bec}

The brightest sources in the catalogues have \hbox{$0.5$--$8\kev$} fluxes up to \hbox{$\sim(1$--$3)\times10^{-13}\ergpcmsqps$}. There is a non-negligible contribution to the whole-sky background by sources brighter than this; in fact, around \hbox{$15$--$20$} per cent of the total XRB is due to bright, rare sources that are not sampled in pencil-beam surveys. A `bright-end correction' is necessary to recover this missing intensity.

In order to remove some of the bright-end variation we truncate the source lists above \hbox{$5\times10^{-14}\ergpcmsqps$} (\hbox{$0.5$--$8\kev$} band). This is the same truncation as used in Fig.~\ref{raw_fractions}(b) and corresponds to removing sources that only occur with densities of \hbox{$\sim$} a few in the \hbox{$\sim0.1\degsq$} CDF and XMM-LH fields. With the bright-end sources removed the expected missing flux can be estimated by integrating known \hbox{$\log N$--$\log S$} distributions. These are available in the broad \hbox{$0.5$--$2$} and \hbox{$2$--$10\kev$} bands but not in the narrow bands used in this work. In order to calculate the contribution in a narrow band it is necessary to assume the spectral shape of the sources. Here we assume a power-law spectrum but the typical spectral slope of such a power-law can be a strong function of source flux.

A number of studies point to significant evolution of the spectral shape with flux \citep{ueda99,fiore03,alexander03,streblyanska03}, with sources brighter than \hbox{$\sim10^{-14}\ergpcmsqps$} having typical spectral slopes much softer than the \hbox{$\Gamma=1.4$} of the total background. Shallow surveys \citep[see e.g.]{ueda99} find that the brightest sources, with X-ray fluxes \hbox{$\gtrsim10^{-13}\ergpcmsqps$}, have spectral slopes of \hbox{$\Gamma\sim1.8$--$2.2$}. \citet{streblyanska03} fitted a power-law spectrum to the stacked spectra of sources of a \hbox{$500\ks$} \xmm\ observation in the CDF-S. Fitting the hard band (\hbox{$1$--$8\kev$}) with a power-law reveals a strong correlation between spectral slope and flux, with \hbox{$\Gamma\sim1.8$} for \hbox{$\sim10^{-13}\ergpcmsqps$} (\hbox{$2$--$10\kev$} flux) decreasing to \hbox{$\Gamma\sim1.5$} at \hbox{$\sim10^{-14}\ergpcmsqps$} and \hbox{$\Gamma\sim1.2$} at \hbox{$\sim10^{-15}\ergpcmsqps$}. \citet{alexander03} see sources with \hbox{$\Gamma\sim1$} at fluxes \hbox{$\lesssim10^{-16}\ergpcmsqps$} in the CDF-N.

We haven taken a linear fit to the evolution of spectral index with flux based on the relationship as observed by \citet{streblyanska03}, although we conservatively impose a maximum value of \hbox{$\Gamma=2$}. Such a model is only an approximation to the actual dependence of spectral shape with flux. It does, however, follow the trends observed by \citet{streblyanska03} in the CDF-S (using both \xmm\ and \chandra\ observations) and \citet{ueda99} using the ASCA Medium-Sensitivity Survey, and whilst crude, it is important to take into account the fact that the bright-end correction is due to sources which are significantly softer than those which we have already stacked. Given our model, the bright-source contribution in each of the \hbox{$2$--$4$}, \hbox{$4$--$6$}, \hbox{$6$--$8$} and \hbox{$8$--$12\kev$} narrow bands can then be calculated, using an appropriate spectral index at each source flux, from the \hbox{$2$--$10\kev$} flux contribution obtained by integrating the \citet{moretti03} hard band \hbox{$\log N$--$\log S$} distribution (the most complete published to date). 

Correcting the soft bands (\hbox{$0.2$--$0.5$}, \hbox{$0.5$--$1$} and \hbox{$1$--$2\kev$}) is more complex, since the shape of the background is not well constrained. The source stacking approach of \citet{streblyanska03} in the \hbox{$0.4$--$8\kev$} range revealed a systematically softer spectrum than that seen for \hbox{$1$--$8\kev$} (an increase in \hbox{$\Gamma$} of \hbox{$\sim0.1$}). No studies cover the \hbox{$0.5$--$2\kev$} range ideally required for the soft band bright-end correction so we take the same linear fit to the evolution of spectral index as used in the hard band, but with an increase in \hbox{$\Gamma$} by 0.1. We use the \citet{moretti03} \hbox{$0.5$--$2\kev$} soft band \hbox{$\log N$--$\log S$} distribution. There is now some evidence to suggest that the \citet{moretti03} \hbox{$\log N$--$\log S$} slope is too steep at the bright end compared to results from the RIXOS, RBS and NEPS surveys \citep{mason00,schwope00,gioia03}; thus, our bright-end correction in the soft band could potentially be an underestimate.

\begin{figure}
\rotatebox{270}{
\resizebox{!}{\columnwidth}
{\includegraphics{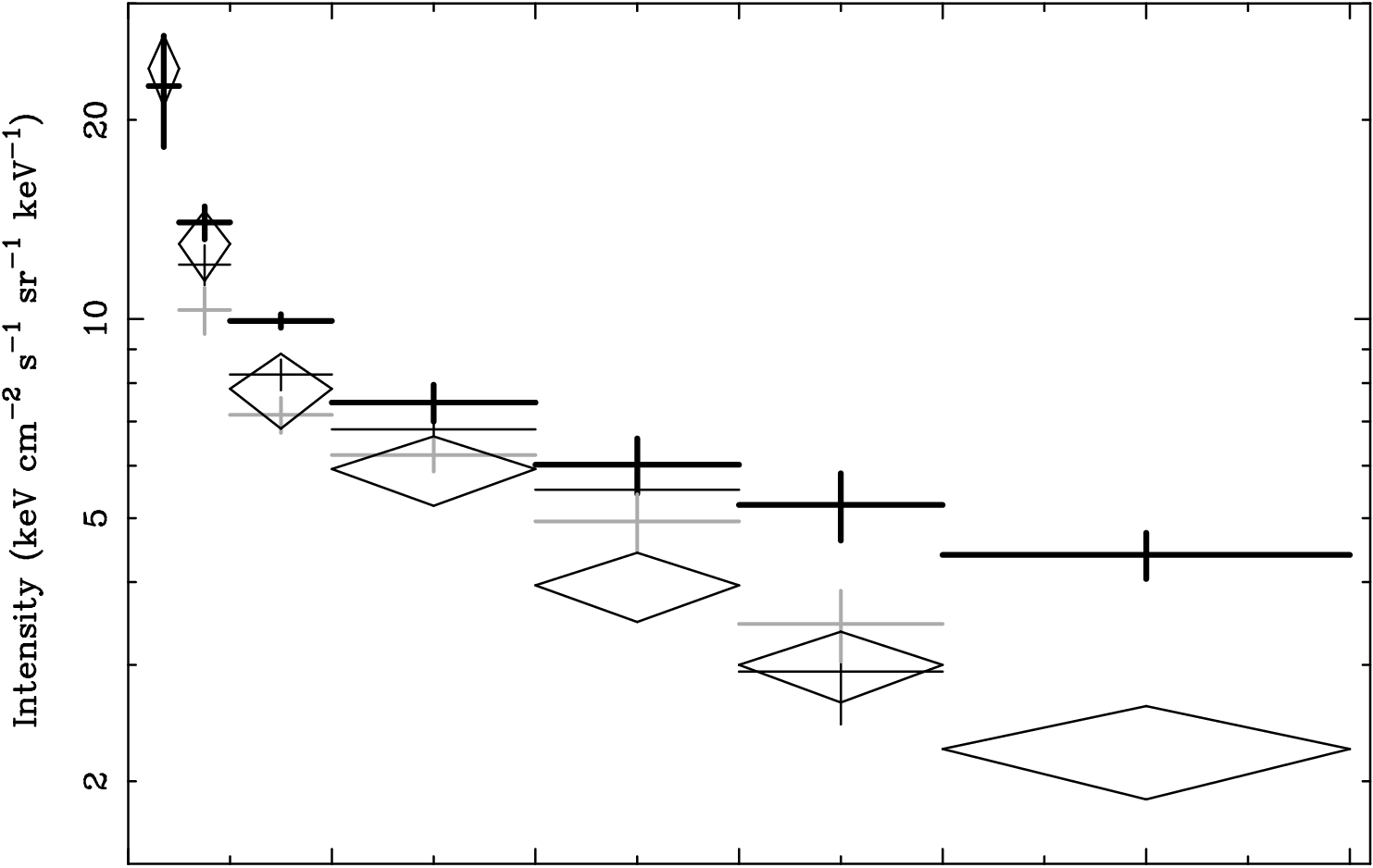}}
}
\rotatebox{270}{
\resizebox{!}{\columnwidth}
{\includegraphics{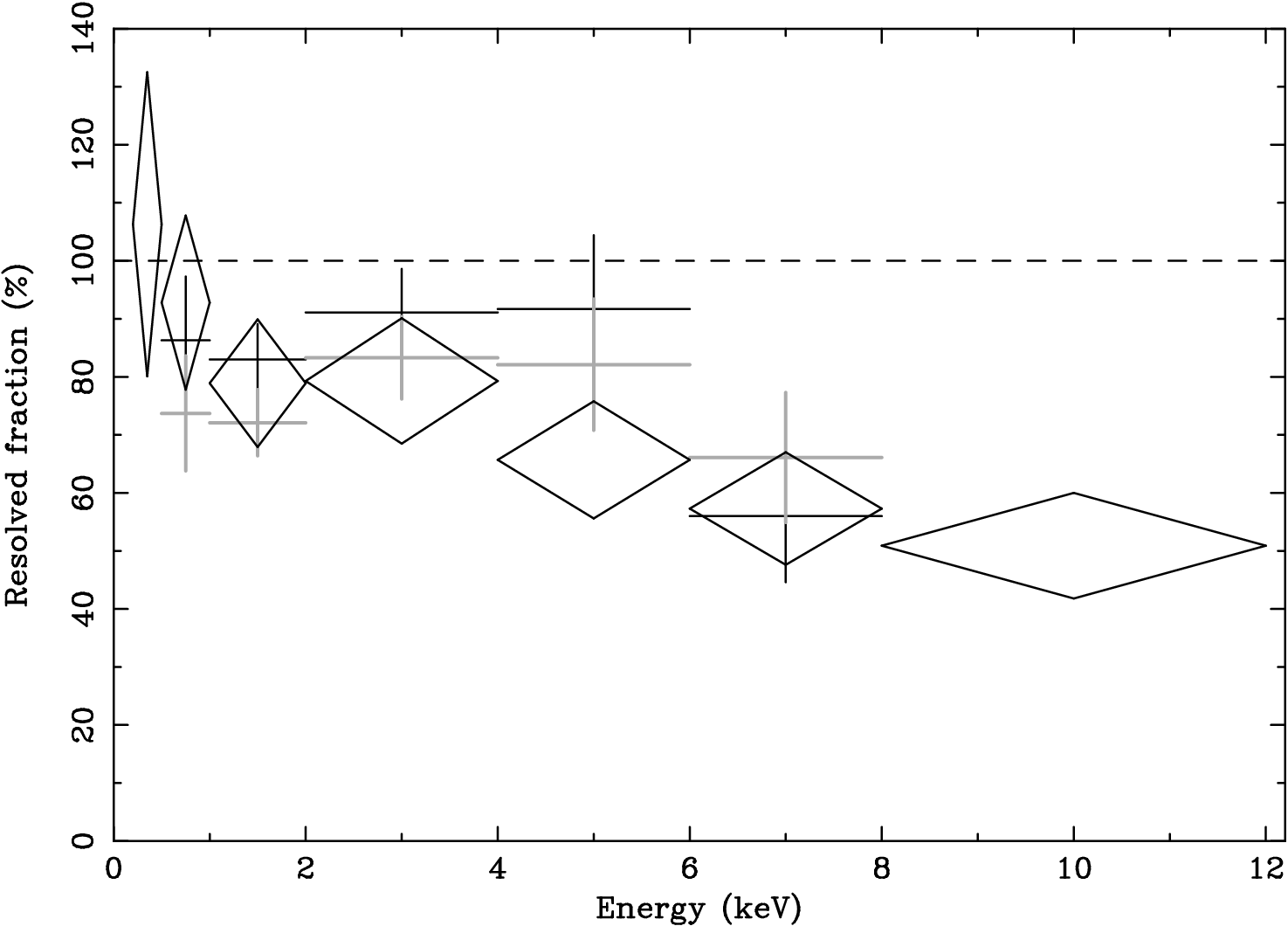}}
}
\caption{The upper panel shows the total extragalactic XRB (thick black crosses) along with the intensity resolved into sources in the CDF-N (black crosses), CDF-S (grey crosses) and combined PN/MOS-1/MOS-2 results in the XMM-LH (black diamonds). The resolved intensities have been corrected for the bright-end population by removing all sources with \hbox{$0.5$--$8\kev$} flux \hbox{$>5\times10^{-14}\ergpcmsqps$} and then adding the missing flux from sources brighter than this by integrating \hbox{$\log N$--$\log S$} distributions (refer to section \ref{bec}). The lower panel shows the resolved intensities as fractions of the total extragalactic XRB in each energy band.}
\label{corrected_fractions}
\end{figure}

Fig.~\ref{corrected_fractions} shows the bright-end corrected resolved background fractions. The XMM-LH, CDF-S and CDF-N are all consistent with a similar resolved fraction of \hbox{$\sim70$--$90$} per cent up to \hbox{$\sim4\kev$}. The differences are of order \hbox{$\sim10$} per cent which is not inconsistent with what would be expected from cosmic variance between the fields \citep[see][]{barcons00}. \xmm\ sees a drop in the resolved fraction in the \hbox{$4$--$6\kev$} band whereas the CDFs are still consistent with a high resolved fraction of \hbox{$\sim80$--$90$} per cent. This difference is due to the faint, hard sources which are not detected in the XMM-LH; if these sources are removed from the \chandra\ data then all three data-sets show agreement in the \hbox{$4$--$6\kev$} fraction. Most importantly, the picture is different at high energies; \xmm\ and both \chandra\ deep fields only resolve \hbox{$\sim50$--$70$} per cent of the background in the \hbox{$6$--$8\kev$} band, much less than the fraction at lower energies. The \hbox{$8$--$12\kev$} band, as probed by \xmm, is only \hbox{$\sim40$--$60$} per cent resolved. 

We find the total resolved fluxes in the \hbox{$0.5$--$2$} and \hbox{$2$--$8\kev$} bands to be \hbox{$(6.9\pm0.2)\times10^{-12}$} and \hbox{$(1.5\pm0.1)\times10^{-11}\ergpcmsqps$} for the CDF-N. These compare well with the \hbox{$(6.7\pm0.3)\times10^{-12}$} and \hbox{$(1.66\pm0.05)\times10^{-11}\ergpcmsqps$} reported in \citet{bauer04} from a detailed analysis of the combined CDF number counts. The fluxes can also be compared to the \hbox{$(7.1\pm0.3)\times10^{-12}$} and \hbox{$(1.43\pm0.08)\times10^{-11}\ergpcmsqps$} determined integrating the \hbox{$\log N$--$\log S$} curves of \citet{moretti03} (although note that these do not contain the \hbox{$2\ms$} extension to the original \hbox{$1\ms$} CDF-N).


We find the total \hbox{$0.5$--$2$} and \hbox{$2$--$8\kev$} resolved fractions are \hbox{$\sim85$} and \hbox{$\sim80$} per cent for the CDF-N (using the total XRB spectrum we describe in section \ref{resolving}). Such high fractions are often used to claim that the XRB is almost completely resolved into discrete sources even in the broad \hbox{$2$--$10\kev$} band. Our results suggest that stating a resolved fraction in this manner is misleading, since it is the high resolved fractions in the \hbox{$2$--$6\kev$} range which are dominating the total, whilst the resolved fraction beyond \hbox{$6\kev$} is significantly lower. 

\section{Discussion}

Three independent observations now point to a significant reduction in the resolved fraction of the X-ray background at the hardest energies. \xmm\ starts to see a downturn above \hbox{$\sim4\kev$} whilst \chandra\ is sensitive to the fainter, harder sources, which fill in some of the missing fraction and push the downturn to energies \hbox{$\gtrsim6\kev$}. The missing fraction has the spectral signature of highly obscured AGN and could be evidence of the large population of heavily absorbed objects which are thought to account for the \hbox{$30\kev$} peak of XRB intensity \citep[see e.g.][]{maiolino98,maiolino03}. 

To quantify the possible nature of an undetected population, we modelled the spectral shape of the unresolved background as a function of redshift and intrinsic absorption column density. A grid of spectra was built in the range \hbox{$z=0.1$--$3$} and \hbox{$N_{\rm{H}}=10^{22}$--$10^{25}\pcmsq$}. \textsc{xspec} \textsc{v11.3.1} was used to generate the spectra using a \textsc{pexrav} model \citep{magdziarz95}, with a \hbox{$\Gamma=2$} power-law and an \hbox{$R=1$} reflection component  (see e.g. \citealp{malizia03}), multiplied by \textsc{zwabs} \citep{morrison83} photoelectric absorption at the source redshift. 

\begin{figure}
\rotatebox{270}{
\resizebox{!}{\columnwidth}
{\includegraphics{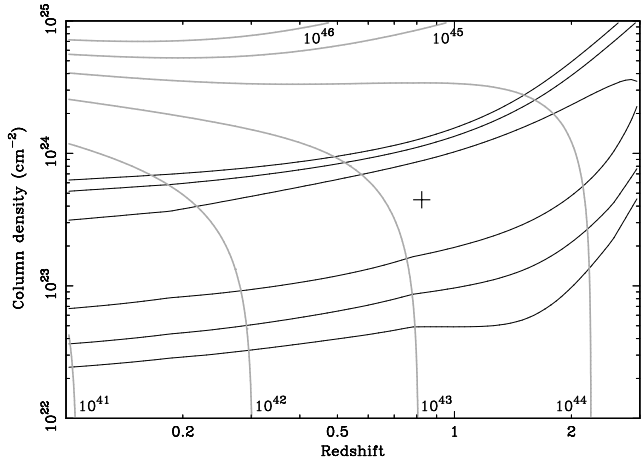}}}
\caption{Contour plot of fits to the spectral shape of the missing XRB spectrum for the XMM-LH data in the $2$--$12\kev$ range. $68$, $90$ and $95$ per cent confidence contours are shown. The best-fitting point is indicated at \hbox{$z\sim0.8$} and \hbox{$N_{\rm{H}}\sim4.5\times10^{23}\pcmsq$}. The grey contours indicate the maximum, unobscured, rest-frame \hbox{$2$--$10\kev$} luminosity (\hbox{$\rm erg~s^{-1}$}) of the source population required such that the sources remain below the sensitivity limit (estimated at \hbox{$\sim1.5\times10^{-15}\ergpcmsqps$} in a \hbox{$4$--$8\kev$} band).}
\label{obs_population}
\end{figure}

We computed the goodness-of-fit between each model spectrum and the residual XRB spectrum (i.e. the difference between the resolved and total backgrounds). Fig.~\ref{obs_population} shows the confidence contours in the redshift and absorption column density required to account for the shape of the missing AGN population for the XMM-LH. This is a rather simplistic approximation to the spectra of highly obscured sources; specifically, we neglect the effects of iron K emission and scattered-flux components, as well as assuming that the missing sources occur at a fixed redshift and with a fixed absorption column density. Nevertheless, some generalised conclusions about the nature of the unresolved population can be drawn.

\begin{figure}
\rotatebox{270}{
\resizebox{!}{\columnwidth}
{\includegraphics{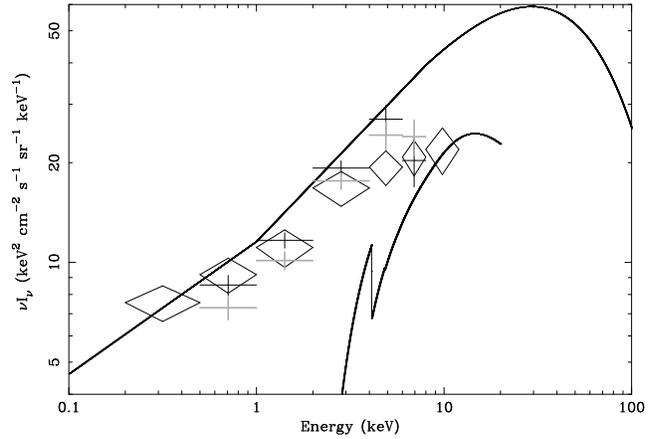}}}
\caption{The upper curve shows the total extragalactic XRB (as described in section \ref{resolving}) in \hbox{$\nu I_{\nu}$}. The intensity resolved into sources in the CDF-N, CDF-S and XMM-LH are shown as black crosses, grey crosses and black diamonds respectively. The lower curve shows the spectrum of the best-fitting model for the missing AGN population with \hbox{$z=0.8$} and \hbox{$N_{\rm{H}}=4.5\times10^{23}\pcmsq$}. Sharp features (such as that due to iron at \hbox{$4\kev$}), is a consequence of our assumption that the missing population is composed entirely of AGN with this spectrum. In reality the missing population will be mixture of sources at different redshifts and with different levels of absorption; any sharp features will be smeared-out.}
\label{nu_i_nu}
\end{figure}

The XMM-LH, CDF-N and CDF-S all show similar contour diagrams for the characteristics of the missing source population. The XMM-LH data, which extends to the \hbox{$8$--$12\kev$} band, shows the most constrained set of contours although the degeneracy between low-z/low-$N_{\rm{H}}$ and high-z/high-$N_{\rm{H}}$ is clear. The best-fitting point is at \hbox{$z\sim0.8$} and \hbox{$N_{\rm{H}}\sim4.5\times10^{23}\pcmsq$}. For the CDF-N and CDF-S the lack of data above \hbox{$8\kev$} results in much more relaxed contours although there is again some preference for sources at \hbox{$z\sim0.5$--$1.5$} with \hbox{$N_{\rm{H}}\sim10^{23}$--$10^{24}\pcmsq$}. Figure \ref{nu_i_nu} shows the total and resolved levels of the XRB in \hbox{$\nu I_{\nu}$} as well as the intensity that would be contributed by the best-fitting model of an obscured population.

We conclude that whilst the data are unable to constrain tightly the spectral shape of the missing XRB sources, the most plausible population would seem be of objects at redshifts of \hbox{$\sim0.5$--$1.5$}, with unabsorbed luminosities of \hbox{$\lesssim5\times10^{43}\ergps$}, but heavily obscured by column densities of \hbox{$\sim10^{23}$--$10^{24}\pcmsq$}. The CDFs probe deeper than the XMM-LH but still fail to detect the missing sources in the \hbox{$2$--$4\kev$} band, or our stacking analysis would have revealed the higher energy emission. Any soft starburst or scattered emission in these objects must be small and this may suggest a high covering fraction for the absorption.

In order to provide the missing fraction of the hard X-ray background these sources must have a density on the sky of \hbox{$\gtrsim2800\pdegsq$}. This would correspond to a lower limit of around 350 undetected AGN in the CDF-N. There must therefore be \hbox{$>3$} times more undetected, obscured sources than detected, unobscured ones, in the \hbox{$4$--$8\kev$} band. This requires that the sources in the new population have a high covering fraction of obscuring matter. 

From the absorption-corrected flux we can estimate the black hole mass density of this obscured population to be \hbox{$(6$--$9\times10^{4})~\epsilon^{-1}_{-1}~\kappa^{}_{-1}\msunpmpccub$}; where \hbox{$\epsilon_{-1}$} is the accretion efficiency in units of 0.1 and \hbox{$\kappa^{}_{-1}$} is a bolometric correction in units of 0.1. We have used a similar calculation to \citet{fabian99} although note that the bolometric correction is \hbox{$\sim10$--$15$} for our lower-luminosity \citep{fabian04}. The missing population thus accounts for \hbox{$\sim10$--$20$} per cent of the local black hole mass density of \hbox{$4$--$5\times10^{5}\msunpmpccub$} \citep[see e.g.][]{shankar04}, the rest of which is due to bright AGN and, in particular, quasars, where the bolometric correction is \hbox{$\sim30$} or more, and where a typical redshift of \hbox{$z\sim2$} is more appropriate (refer to \citealt{fabian99} and references therein).

The steepness of the \hbox{$5$--$10\kev$} source number counts, with no evidence for a break to a flatter slope at the faint end, also indicates that there remains a substantial number of objects which have yet to be found in this band \citep{hasinger01,rosati02,baldi02}. Recent optical and infra-red data from the HST and Spitzer telescopes, acquired as part of the GOODS project, are also providing evidence for large numbers of obscured AGN at \hbox{$z>1$} \citep[e.g.][]{treister04}. Deep, multi-wavelength efforts such as GOODS may well be the best way to identify the missing population; confirmation of Compton-thick objects will ultimately require X-ray missions with high sensitivities beyond $10\kev$, such as the next generation of focusing hard X-ray telescopes.

\section{Conclusions}

\begin{itemize}
\item 
Whilst the XRB is \hbox{$\sim85$} and \hbox{$\sim80$} per cent resolved in the broad \hbox{$0.5$--$2$} and \hbox{$2$--$10\kev$} bands respectively (depending on XRB normalisation), it is only \hbox{$\sim60$} per cent resolved above \hbox{$\sim6\kev$} and \hbox{$\sim50$} per cent resolved above \hbox{$\sim8\kev$}. This decrease in resolved fraction is seen in both the CDFs and the XMM-LH.
\item
The decrease in resolved fraction as seen in XMM-LH occurs at energies \hbox{$\gtrsim4\kev$} whilst in the CDFs the drop occurs at \hbox{$\gtrsim6\kev$}. This difference is due to faint, hard sources which are detected in CDFs but not the XMM-LH. All three surveys are unable to account for almost half of the X-ray background above \hbox{$\sim6\kev$}.
\item
The missing fraction has a spectral shape that is consistent with that which would be expected from a population of faint, heavily obscured AGN located at a redshift of \hbox{$\sim0.5$--$1.5$} and with intrinsic absorption columns of \hbox{$\sim10^{23}$--$10^{24}\pcmsq$}.
\item
The deep CDF data show that any such obscured population is not detected in the \hbox{$2$--$4\kev$} band, or our stacking analysis would have revealed the higher energy emission; consequently, any soft scattered or starburst emission from these objects must be small. 
\end{itemize}

\section{Acknowledgments}

Based on observations with \xmm, an ESA science mission with instruments and contributions directly funded by ESA Member States and the USA (NASA). MAW and FEB acknowledge support from PPARC. ACF and DMA thank the Royal Society for support. WNB thanks the CXC grant GO2-31287A. We would also like to thank the anonymous referee for help comments.


\bibliographystyle{mnras} 
\bibliography{mn-jour,worsley_9dec04.bib}

\begin{thebibliography}{}

\bibitem[\protect\citeauthoryear{{Alexander} et~al.}{{Alexander}
  et~al.}{2003}]{alexander03}
{Alexander} D.~M. et~al., 2003, \aj, 126, 539

\bibitem[\protect\citeauthoryear{{Baldi} et~al.}{{Baldi}
  et~al.}{2002}]{baldi02}
{Baldi} A., {Molendi} S., {Comastri} A., {Fiore} F., {Matt} G.,  {Vignali} C.,
  2002, \apj, 564, 190

\bibitem[\protect\citeauthoryear{{Barcons}, {Mateos}, \& {Ceballos}}{{Barcons}
  et~al.}{2000}]{barcons00}
{Barcons} X., {Mateos} S.,  {Ceballos} M.~T., 2000, \mnras, 316, L13

\bibitem[\protect\citeauthoryear{{Bauer} et~al.}{{Bauer}
  et~al.}{2004}]{bauer04}
{Bauer} F.~E., {Alexander} D.~M., {Brandt} W.~N., {Schneider} D.~P., {Treister}
  E., {Hornschemeier} A.~E.,  {Garmire} G.~P., 2004, \aj, 128, 2048

\bibitem[\protect\citeauthoryear{{Campana} et~al.}{{Campana}
  et~al.}{2001}]{campana01}
{Campana} S., {Moretti} A., {Lazzati} D.,  {Tagliaferri} G., 2001, \apjl, 560,
  L19

\bibitem[\protect\citeauthoryear{{Comastri}}{{Comastri}}{2004}]{comastri04}
{Comastri} A., 2004 (astro-ph/0406031)

\bibitem[\protect\citeauthoryear{{De Luca} \& {Molendi}}{{De Luca} \&
  {Molendi}}{2004}]{deluca04}
{De Luca} A.,  {Molendi} S., 2004, \aap, 419, 837

\bibitem[\protect\citeauthoryear{{Fabian}}{{Fabian}}{2004}]{fabian04}
{Fabian} A.~C., 2004, in Coevolution of Black Holes and Galaxies, p. 447

\bibitem[\protect\citeauthoryear{{Fabian} \& {Iwasawa}}{{Fabian} \&
  {Iwasawa}}{1999}]{fabian99}
{Fabian} A.~C.,  {Iwasawa} K., 1999, \mnras, 303, L34

\bibitem[\protect\citeauthoryear{{Fiore} et~al.}{{Fiore}
  et~al.}{2003}]{fiore03}
{Fiore} F. et~al., 2003, \aap, 409, 79

\bibitem[\protect\citeauthoryear{{Franceschini}, {Braito}, \&
  {Fadda}}{{Franceschini} et~al.}{2002}]{franceschini02}
{Franceschini} A., {Braito} V.,  {Fadda} D., 2002, \mnras, 335, L51

\bibitem[\protect\citeauthoryear{{Gandhi} \& {Fabian}}{{Gandhi} \&
  {Fabian}}{2003}]{gandhi03}
{Gandhi} P.,  {Fabian} A.~C., 2003, \mnras, 339, 1095

\bibitem[\protect\citeauthoryear{{Giacconi} et~al.}{{Giacconi}
  et~al.}{1962}]{giacconi62}
{Giacconi} R., {Gursky} H., {Paolini} F.~R.,  {Rossi} B.~B., 1962, Physical
  Review Letters, 9, 439

\bibitem[\protect\citeauthoryear{{Giacconi} et~al.}{{Giacconi}
  et~al.}{2002}]{giacconi02}
{Giacconi} R. et~al., 2002, \apjs, 139, 369

\bibitem[\protect\citeauthoryear{{Gilli}, {Salvati}, \& {Hasinger}}{{Gilli}
  et~al.}{2001}]{gilli01}
{Gilli} R., {Salvati} M.,  {Hasinger} G., 2001, \aap, 366, 407

\bibitem[\protect\citeauthoryear{{Gioia} et~al.}{{Gioia}
  et~al.}{2003}]{gioia03}
{Gioia} I.~M., {Henry} J.~P., {Mullis} C.~R., {B{\" o}hringer} H., {Briel}
  U.~G., {Voges} W.,  {Huchra} J.~P., 2003, \apjs, 149, 29

\bibitem[\protect\citeauthoryear{{Gruber} et~al.}{{Gruber}
  et~al.}{1999}]{gruber99}
{Gruber} D.~E., {Matteson} J.~L., {Peterson} L.~E.,  {Jung} G.~V., 1999, \apj,
  520, 124

\bibitem[\protect\citeauthoryear{{Hasinger} et~al.}{{Hasinger}
  et~al.}{2001}]{hasinger01}
{Hasinger} G. et~al., 2001, \aap, 365, L45

\bibitem[\protect\citeauthoryear{{Hasinger} et~al.}{{Hasinger}
  et~al.}{1998}]{hasinger98}
{Hasinger} G., {Burg} R., {Giacconi} R., {Schmidt} M., {Trumper} J.,
  {Zamorani} G., 1998, \aap, 329, 482

\bibitem[\protect\citeauthoryear{{Lockman}}{{Lockman}}{2003}]{lockman04}
{Lockman} F.~J., 2003 (astro-ph/0311386)

\bibitem[\protect\citeauthoryear{{Lockman}, {Jahoda}, \& {McCammon}}{{Lockman}
  et~al.}{1986}]{lockman86}
{Lockman} F.~J., {Jahoda} K.,  {McCammon} D., 1986, \apj, 302, 432

\bibitem[\protect\citeauthoryear{{Magdziarz} \& {Zdziarski}}{{Magdziarz} \&
  {Zdziarski}}{1995}]{magdziarz95}
{Magdziarz} P.,  {Zdziarski} A.~A., 1995, \mnras, 273, 837

\bibitem[\protect\citeauthoryear{{Maiolino} et~al.}{{Maiolino}
  et~al.}{2003}]{maiolino03}
{Maiolino} R. et~al., 2003, \mnras, 344, L59

\bibitem[\protect\citeauthoryear{{Maiolino} et~al.}{{Maiolino}
  et~al.}{1998}]{maiolino98}
{Maiolino} R., {Salvati} M., {Bassani} L., {Dadina} M., {della Ceca} R., {Matt}
  G., {Risaliti} G.,  {Zamorani} G., 1998, \aap, 338, 781

\bibitem[\protect\citeauthoryear{{Malizia} et~al.}{{Malizia}
  et~al.}{2003}]{malizia03}
{Malizia} A., {Bassani} L., {Stephen} J.~B., {Di Cocco} G., {Fiore} F.,  {Dean}
  A.~J., 2003, \apjl, 589, L17

\bibitem[\protect\citeauthoryear{{Marshall} et~al.}{{Marshall}
  et~al.}{1980}]{marshall80}
{Marshall} F.~E., {Boldt} E.~A., {Holt} S.~S., {Miller} R.~B., {Mushotzky}
  R.~F., {Rose} L.~A., {Rothschild} R.~E.,  {Serlemitsos} P.~J., 1980, \apj,
  235, 4

\bibitem[\protect\citeauthoryear{{Mason} et~al.}{{Mason}
  et~al.}{2000}]{mason00}
{Mason} K.~O. et~al., 2000, \mnras, 311, 456

\bibitem[\protect\citeauthoryear{{Moretti} et~al.}{{Moretti}
  et~al.}{2003}]{moretti03}
{Moretti} A., {Campana} S., {Lazzati} D.,  {Tagliaferri} G., 2003, \apj, 588,
  696

\bibitem[\protect\citeauthoryear{{Moretti} et~al.}{{Moretti}
  et~al.}{2002}]{moretti02}
{Moretti} A., {Lazzati} D., {Campana} S.,  {Tagliaferri} G., 2002, \apj, 570,
  502

\bibitem[\protect\citeauthoryear{{Morrison} \& {McCammon}}{{Morrison} \&
  {McCammon}}{1983}]{morrison83}
{Morrison} R.,  {McCammon} D., 1983, \apj, 270, 119

\bibitem[\protect\citeauthoryear{{Roberts} \& {Warwick}}{{Roberts} \&
  {Warwick}}{2001}]{roberts01}
{Roberts} T.~P.,  {Warwick} R.~S., 2001, in ASP Conf. Ser. 234: X-ray Astronomy
  2000, p. 569

\bibitem[\protect\citeauthoryear{{Rosati} et~al.}{{Rosati}
  et~al.}{2002}]{rosati02}
{Rosati} P. et~al., 2002, \apj, 566, 667

\bibitem[\protect\citeauthoryear{{Schwope} et~al.}{{Schwope}
  et~al.}{2000}]{schwope00}
{Schwope} A. et~al., 2000, Astronomische Nachrichten, 321, 1

\bibitem[\protect\citeauthoryear{{Shankar} et~al.}{{Shankar}
  et~al.}{2004}]{shankar04}
{Shankar} F., {Salucci} P., {Granato} G.~L., {De Zotti} G.,  {Danese} L., 2004,
  \mnras, 387

\bibitem[\protect\citeauthoryear{{Stark} et~al.}{{Stark}
  et~al.}{1992}]{stark92}
{Stark} A.~A., {Gammie} C.~F., {Wilson} R.~W., {Bally} J., {Linke} R.~A.,
  {Heiles} C.,  {Hurwitz} M., 1992, \apjs, 79, 77

\bibitem[\protect\citeauthoryear{{Streblyanska} et~al.}{{Streblyanska}
  et~al.}{2003}]{streblyanska03}
{Streblyanska} A., {Bergeron} J., {Brunner} H., {Finoguenov} A., {Hasinger} G.,
   {Mainieri} V., 2003 (astro-ph/0309089)

\bibitem[\protect\citeauthoryear{{Treister} et~al.}{{Treister}
  et~al.}{2004}]{treister04}
{Treister} E. et~al., 2004, \apj, 616, 123

\bibitem[\protect\citeauthoryear{{Ueda} et~al.}{{Ueda} et~al.}{2003}]{ueda03}
{Ueda} Y., {Akiyama} M., {Ohta} K.,  {Miyaji} T., 2003, \apj, 598, 886

\bibitem[\protect\citeauthoryear{{Ueda} et~al.}{{Ueda} et~al.}{1999}]{ueda99}
{Ueda} Y., {Takahashi} T., {Ishisaki} Y., {Ohashi} T.,  {Makishima} K., 1999,
  \apjl, 524, L11

\bibitem[\protect\citeauthoryear{{Worsley} et~al.}{{Worsley}
  et~al.}{2004}]{worsley04}
{Worsley} M.~A., {Fabian} A.~C., {Barcons} X., {Mateos} S., {Hasinger} G.,
  {Brunner} H., 2004, \mnras, 352, L28

\end{thebibliography}

\end{document}